\documentclass[preprint,amsmath,amssymb,aps,nofootinbib]{revtex4}

\usepackage{graphicx}
\usepackage{subfigure}

\newcommand{\wt}{\widetilde}
\newcommand{\ovl}{\overline}
\newcommand{\al}{\alpha}
\newcommand{\be}{\beta}
\newcommand{\la}{\lambda}

\newcommand{\beq}{\begin{equation}}
\newcommand{\eeq}{\end{equation}}

\newcommand{\bea}{\begin{eqnarray}}
\newcommand{\eea}{\end{eqnarray}}
\newcommand{\nl}{\nonumber\\}
\renewcommand{\l}{\left}
\renewcommand{\r}{\right}

\begin{document}

\preprint{\begin{tabular}{l}
\hbox to\hsize{hep-ph/0301269 \hfill KIAS-P03010}\\
\end{tabular} }

%
\title{CP violation in $B \to \phi K_S$  decay at large $\tan\be$}

\author{Seungwon Baek}
\affiliation{School of Physics, KIAS, 207-43 Cheongryangri-dong, 
    Seoul 130-722, Korea }

\date{\today}

\thispagestyle{empty}

\begin{abstract}
We consider the chargino contribution to the CP violation in $B \to \phi K_S$ 
decay in the minimal supersymmetric standard model at large $\tan\be$. 
It is shown that the Wilson coefficient $C_{8g}$ of the chromomagnetic 
penguin operator can be significantly enhanced by the 
chargino-mediated diagrams
while satisfying other direct/indirect experimental constraints.
The enhanced $C_{8g}$ allows large deviation in the CP asymmetry
from the standard model prediction,
especially it can explain the apparent anomaly reported by BaBar and Belle.
\end{abstract}


\maketitle


The measurement of CP asymmetries at B-factories is a powerful
probe of new physics (NP). 
In the standard model (SM), the origin of CP violation is the
phase $\delta_{\rm CKM}$ in the $3\times 3$ Cabibbo-Kobayashi-Maskawa (CKM) 
matrix of quark mixing. In general, there can be many new sources of
CP violation in new physics models beyond the SM.

The time-dependent CP asymmetry in the neutral $B$ decays to CP eigenstates
$B \to f_{\rm CP}$
gives information on the two classes of CP violation $C_f$ and $S_f$~\cite{Nir:2002gu}:
\bea
  A_{\rm CP}(t) &=&\frac{\Gamma(\ovl{B}(t)\to f_{\rm CP})-\Gamma({B}(t)\to f_{\rm CP})}
  {\Gamma(\ovl{B}(t)\to f_{\rm CP})+\Gamma({B}(t)\to f_{\rm CP})} \nl
  &=& -C_f \cos (\Delta m_B t) + S_f \sin (\Delta m_B t),
\eea
where $\Delta m_B$ is the mass difference of the $B$ system and
$B(\ovl{B})(t)$ is the state at time $t$ which
was $B(\ovl{B})$ at $t=0$. The CP asymmetries $C_f$ and $S_f$
are determined by
\bea
  \la_{\rm CP} \equiv e^{-2i(\be+\theta_d)} \frac{\overline{A}}{A},
\eea
where $\be (\theta_d)$ is the contribution of SM (NP) to 
the phase in the $B-\ovl{B}$ mixing
and $\ovl{A}(A)$ is the decay amplitude for  $\ovl{B}(B)\to f_{\rm CP}$.

The time-dependent CP asymmetry in $B\to J/\psi K$ decay
measured at BaBar~\cite{sin2b_babar} and Belle~\cite{sin2b_belle}
shows the existence of CP violation in the $B$-meson system, and the current
world average~\cite{Nir:2002gu}
\bea
 \sin 2\be_{J/\psi K_S} \equiv S_{J/\psi K_S} = 0.734 \pm 0.054.
\label{eq:sin2b_exp}
\eea
is fully consistent with the CKM picture of CP violation of the SM.
In the SM, the CP asymmetry in $B\to \phi K_S$ also measures the same angle 
angle $\be$ because the CP violation in the decay amplitude 
is suppressed by the Cabibbo angle $\la$ ($\approx 0.22$):
$S_{\phi K_S} = S_{J/\psi K_S}+{\cal O}(\lambda^2)$.

Recently, BaBar~\cite{b2kp_babar} and Belle~\cite{b2kp_belle}
have announced the first measurement of time-dependent
CP asymmetries in $B\to \phi K_S$. The weighted average of the mixing-induced
CP asymmetry
\bea
 \sin 2\be_{\phi K_S} \equiv S_{\phi K_S} = -0.39 \pm 0.41
\label{eq:sin_b2kp}
\eea
differs from the SM prediction of (\ref{eq:sin2b_exp}) by 2.7$\sigma$.
The Belle collaboration has also reported the direct CP asymmetry
\bea
C_{\phi K} = 0.56 \pm 0.43,
\label{eq:cos_b2kp}
\eea
which is consistent with zero as predicted in the SM.

Contrary to the $B\to J/\psi K_S$ decay where $\overline{A}/A=1$ to a
good approximation even in the presence of NP,
the loop-induced process $B \to \phi K_S$ generally allows large deviation
from 1 in phase and modulus of $\overline{A}/A$.
There are already many works which explain
the apparent deviation (\ref{eq:sin_b2kp}) in various
NP models~\cite{Hiller:2002ci,Ciuchini:2002pd,R-parity,
Raidal:2002ph,b2kp_glu}.

The measured $\sin 2\be$'s from $B \to \eta' K_S$, $
B\to (K^+K^-)K_S$~\cite{b2kp_belle}
and $\sin 2\al$ from $B \to \pi^+\pi^-$~\cite{hep-ex/0301032}
are consistent with the SM although the errors are large.
We note that, however, $B\to \phi K_S$ is unique in that
it doesn't have the tree-level amplitude in the SM unlike 
most other $B$-decays.
Therefore, it is not unlikely
that the NP manifests itself only in the $B\to \phi K_S$ 
decay~\cite{baek_long}.

The minimal supersymmetric(SUSY) standard model (MSSM) 
has many new CP violating phases besides the CKM phase
of the SM. With CP violating phases of order one the 
electric dipole moments (EDMs) easily
exceed the experimental upper bounds by several orders of magnitude.
In addition, the general structure of the sfermion
mass matrices in the generation space leads 
unacceptable flavor changing neutral
currents (FCNC) by gluino mediation. 
These SUSY CP and SUSY FCNC problems strongly constrain the
MSSM parameters.

In the MSSM, it has been shown that (\ref{eq:sin_b2kp}) can be accomodated
if there exist new flavor structures in the
up- or down-type squark mass 
matrices~\cite{Hiller:2002ci,b2kp:before,b2kp_glu}.
In this paper, we will show that the chargino contribution
can generate large deviation in $S_{\phi K}$ at large $\tan\beta$, 
even if CKM is the only source of flavor mixing.

Specifically, we adopt a decoupling scenario where the masses of
the first two generation scalar fermions are very heavy 
($\gtrsim {\cal O}(10 \mbox{ TeV})$),
so that the SUSY FCNC and SUSY CP problems are solved
without a naturalness problem~\cite{Cohen:1996vb}.
We also assume the flavor-changing off-diagonal 
elements of scalar fermions are vanishing
to guarantee the absence of the gluino mediated FCNC.
In this case the CKM matrix is the only source of flavor mixing, while
there are new CP violating parameters $\mu,M_2$ in the chargino matrix
\bea
  M_C = \left(
  \begin{array}{cc}
   M_2 &  \sqrt{2} m_W \sin\be \\
  \sqrt{2} m_W \cos\be & \mu
  \end{array}
\right)
\eea
and $A_t$
in the scalar top mass-squared matrix
\bea
  M_{\wt{t}}^2 = \left(
   \begin{array}{cc}
    m^2_{\wt{Q}} + m_t^2 + D_L & m_t (A_t^* -\mu \cot\be) \\
   m_t (A_t -\mu^* \cot\be) &    m^2_{\wt{t}} + m_t^2 + D_R
   \end{array}
  \right),
\eea
where $D_L=(1/2-2/3 \sin^2\theta_W)\cos2\be m_Z^2$
and $D_R=2/3 \sin^2\theta_W\cos2\be m_Z^2$.

In this decoupling scenario, it has been shown that 
the light stop and chargino contributions still allows
large direct CP asymmetry in the radiative
$B$ decay up to $\pm 16$\%~\cite{Baek:1998yn}.
It should be noted that the new contribution to $B-\ovl{B}$ mixing 
is very small and this
model is naturally consistent with 
(\ref{eq:sin2b_exp})~\cite{Baek:1998yn}.

%
%
The decay $B \to \phi K_S$ is described by the $\Delta B =1$ effective 
Hamiltonian~\cite{Beneke:2001ev}. The 
chargino contribution to the Wilson coefficients of QCD penguin
operator is given by
\bea
  C_3^{\wt{\chi}^\pm}
 =-{\alpha_s \over 12 \pi} \sum_{I,k=1}^2 
    {m_W^2 \over m^2_{\wt{\chi}^\pm_I}}
   (\chi^{L}_I)^*_{k2}
   (\chi^{L}_I)_{k3}
   P_2(x_{Ik}),
\label{eq:QCD_penguin}
\eea
where $x_{tH}=m^2_t/m^2_{H^\pm}$, 
$x_{Ik} = m^2_{\wt{\chi}_I^\pm}/m^2_{\wt{t}_k}$
and $P_2(x)$ is the loop-function~\cite{baek_long}.
$(\chi_I^{L(R)})_{kq}$ is the stop($k$)--chargino($I$)--down-quark($q_{L(R)}$) coupling:
\bea
(\chi_I^{L})_{kq} &=& -V_{I1}^* S_{\wt{t}_k \wt{t}_L}
+V_{I2}^* S_{\wt{t}_k \wt{t}_R} \frac{m_t}{\sqrt{2}m_W\sin_\be}, \nl
 (\chi_I^{R})_{kq} &=& U_{I2} S_{\wt{t}_k \wt{t}_L} 
\frac{m_q}{\sqrt{2} m_W \cos\be},
\label{eq:chiLR}
\eea
where $U,V$ ($S$) diagonalize(s) the chargino (stop) mass matrix.
The other Wilson cofficients of QCD penguin operators are 
simply related to $C_3$ by
$C_5=C_3, C_4=C_6=-3 C_3$, where we have neglected 
the box diagram contributions
which are suppressed by $\la$.
In the analysis we decouple the charged Higgs contribution by assuming
$m_{H\pm}=1$ TeV. The effect of $H^\pm$ will be mentioned below.
The expressions for the Wilson coefficients of magnetic operators 
$C_{7\gamma, 8g}^{\chi^\pm}$
can be found, for exmample, in~\cite{bsr,baek_long}.

We note that the $C_{7\gamma(8g)}^{\chi^\pm}$ has the enhancement
factors $m_{\wt{\chi}_I^\pm}/m_b$ by the chirality flip inside the loop
and can dominate the SM contribution.
On the other hand the $C_{3,\cdots, 6}^{\chi^\pm}$ preserve the chirality 
and don't have such enhancement factors. 
In addition, due to the super-GIM mechanism,
the chargino contribution to $C_{3,\cdots, 6}$ are much smaller than the SM
values.
The contribution to chirality flipped operators 
$C_{7\gamma(8g)}^{'\wt{\chi}^\pm}$
are suppressed by $m_s/m_b$.
We neglect the contribution of electroweak penguin operators
which are also expected to be negligible in our scenario. 
Therefore the large
deviation in $\overline{A}/A$ should be generated soley by $C_{8g}$
in this scenario.

To calculte the hadronic matrix elements we use the QCD factorization
method in 
ref.~\cite{Beneke:2001ev}. In this approach it has been demonstrated that
the strong phases of QCD penguin operators cancel out in the 
SM~\cite{Ciuchini:2002pd}.
Since the NP contribution to $C_{3,\cdots, 6}$ are negligible in our scenario,
the strong phase is small as in the SM.
The required $C^{\wt{\chi}^\pm}_{8g}(m_b)$ 
($C^{\rm SM}_{8g}(m_b)\approx -0.147$) 
for large deviation of CP asymmetries
can be estimated from the approximate
numerical expression for $\overline{A}$~\cite{Beneke:2001ev,Ciuchini:2002pd}:
\bea
 \ovl{A} &\propto& 
  \sum_{p=u,c} V_{ps}^* V_{pb} (a_3 + a_4^p + a_5) \nl
  &\approx& -3.9 \times 10^{-4} ( 3.7 e^{0.21 i} + 4.5 C_{8g} ).
\eea

\begin{figure}
\begin{center}
\includegraphics[width=0.45\textwidth]{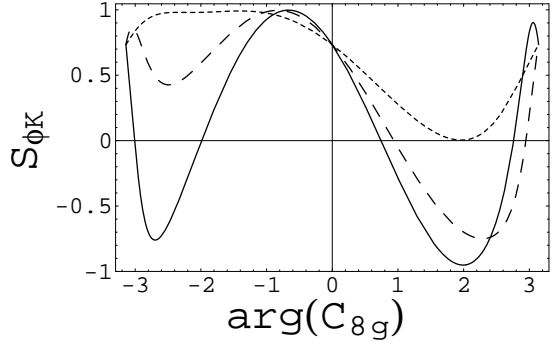}
\end{center}
\caption{$S_{\phi K}$ as a function of $\arg(C_{8g})$ for 
$|C_{8g}|=$0.33 (short dashed line), 0.65 (long dashed line) 
and 1.0 (solid line). }
\label{fig:Skp-C8}
\end{figure}
In Fig.~\ref{fig:Skp-C8}, we show $S_{\phi K}$ as a function of $\arg(C_{8g})$
for $|C_{8g}|=0.33,0.65$ and $1.0$.
From this figure, we can see $|C_{8g}| \approx 0.33-0.65$
with large positive phase can accomodate the deviation within 1$\sigma$.

Since the chargino contributions to $C_{7\gamma}$ and $C_{8g}$
have similar structures, one can think that large deviation in $C_{8g}$ 
may result in the large deviation $C_{7\gamma}$. 
Too large deviation in $C_{7\gamma}$
will violate the measurement of $B(B\to X_s \gamma)$,
for which we take 
$2\times 10^{-4} < B(B\to X_s \gamma) < 4.5 \times 10^{-4}$~\cite{Chen:2001fj},
because it is already consistent with the SM prediction.

Due to the different loop functions, however, the two Wilson coefficients
are not strongly correlated, and it is possible to have large deviation
in $C_{8g}$ while keeping $|C_{7\gamma}|$ constrained to satisfy 
$B(B\to X_s \gamma)$.
Since $\chi^R$ in (\ref{eq:chiLR})
is proportional to $\tan\be$ for large $\tan\be$, 
we need relatively large $\tan\be$
to have sizable effects.

The neutral Higgs boson $h^0$ which is lighter than the $Z$-boson at
tree level has large radiative corrections~\cite{Okada:1990vk} 
which can be approximated at large $\tan\be$ as
\bea
  m_{h^0}^2 \approx m_Z^2 + {3 \over 2 \pi^2} \frac{m_t^4}{v^2}
         \log \l(m_{\wt{t}_1} m_{\wt{t}_2} \over m_t^2\r),
\eea
where $v \approx 246$ GeV.
LEP II sets the lower limit on the SM-like Higgs boson mass:
$m_{h^0} \gtrsim 114.3$ (GeV) at 95\% CL~\cite{Hagiwara:fs}. 
This limit gives very strong constraint on the stop masses.

At large $\tan\be$($\sim {\cal O}(50)$), the SUSY QCD and SUSY electroweak
correction to the non-holomorphic couplings 
$H_u^* D^c Q$ become very important.
This gives large correction on the down-type quark masses
and some CKM matrix elements~\cite{Hall:1993gn}
\bea
  m_b = \frac{\ovl{m}_b}{1+\wt{\epsilon}_3 \tan\be},\quad
  V_{JI} = V_{JI}^{\rm eff} 
  \l[\frac{1+\wt{\epsilon}_3 \tan\be}{1+\epsilon_0 \tan\be}\r]
\eea
where $\wt{\epsilon}_3 \approx \epsilon_0 + \epsilon_Y y_t^2$
and $(JI)=(13)(23)(31)(32)$.
$\ovl{m}_b$, $V_{JI}^{\rm eff}$ are $b$-quark mass and
CKM elements measured at experiments, respectively.
At one-loop and $SU(2)_L\times U(1)_Y$ symmetric limit, the $\epsilon_0$
and $\epsilon_Y$ are given by
\bea
\epsilon_0 &=& {2 \al_s \over 3\pi} {\rm Re} \l(\mu^* \over m_{\wt{g}}\r) 
j(y_{\wt{b}\wt{g}},y_{\wt{Q}\wt{g}}), \nl
\epsilon_Y &=& {1 \over 16\pi^2} {\rm Re} \l(A_t \over \mu\r) 
j(y_{\wt{Q}\mu},y_{\wt{t}\mu}), 
\eea
where $y_{\wt{b}\wt{g}} = m^2_{\wt{b}}/|m_{\wt{g}}|^2$,
$y_{\wt{Q}\wt{g}} = m^2_{\wt{Q}}/|m_{\wt{g}}|^2$,
$y_{\wt{Q}\mu} = m^2_{\wt{Q}}/|\mu|^2$,
$y_{\wt{t}\mu} = m^2_{\wt{t}}/|\mu|^2$. The loop-function is
given by $j(x,y)=(j(x)-j(y))/(x-y)$ with $j(x)=x \log x/(x-1)$.
Note that these SUSY threshold corrections are not
easily decoupled even for very heavy superparticles. 

\begin{figure}
\centering
\includegraphics[width=0.6\textwidth]{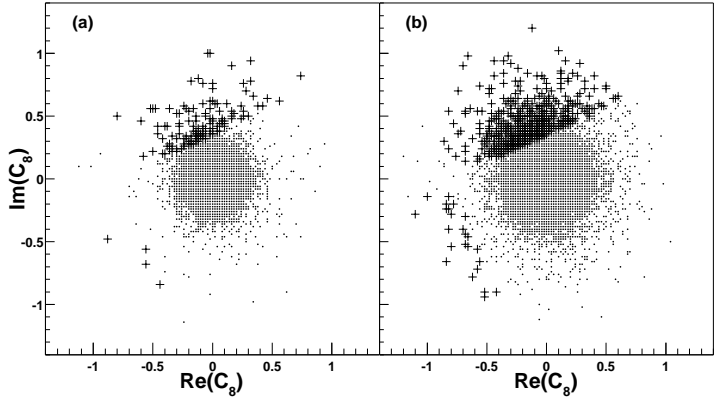}
\caption{Distribution of $C_8$ values for $\tan\be=35$ (a)
and $\tan\be=60$ (b) in the complex plane. 
See the text for other parameters.
The points with $S_{\phi K}<0$ are marked with $+$ symbol.}
\label{fig:C8}
\end{figure}

For the numerical analysis 
we fix the CKM matrix by using 
$|V_{us}|=0.2196$, $|V_{cb}|=4.12\times 10^{-2}$ and 
$|V_{ub}|=3.6\times10^{-3}$~\cite{Hagiwara:fs},
leaving only $\delta_{\rm CKM}$ as free parameter.
For example, for $\delta_{\rm CKM} = \pi/3$, we get 
$\Delta m_{B_d} = 0.491$ ps$^{-1}$ and $\sin2\beta=0.729$,
which are close to the experimental central values~\cite{Hagiwara:fs}.
The free SUSY parameters in our scenario are
$\tan\be$, $M_2$, $\mu$, $m_{\wt{Q}}$, $m_{\wt{t}}$ and $A_t$
($m_{H^\pm}=1$ TeV)
(also $m_{\wt{g}}, m_{\wt{b}}$ are relevant for large $\tan\be$).
In Fig.~\ref{fig:C8}, we show the distribution of $C_{8g}$ in the complex
plane for $\delta_{\rm CKM}=\pi/3$, $\tan\be =35 (60)$, $m_{H^\pm}=1$ TeV,
$ m_{\wt{Q}} = 0.5 $ TeV, $m_{\wt{g}} =1$ TeV, and $m_{\wt{b}_R} = 0.5$ TeV.
We scanned the other parameters as follows: 
\bea
&& 0< m_{\wt{t}} < 1 \mbox{ TeV}, \quad 0 < |\mu| < 1 \mbox{ TeV}, \nl
&& 0< |A_t| < 2 \mbox{ TeV}, \quad 0 < |M_2| < 1 \mbox{ TeV}, \nl
&& -\pi < \arg(\mu),\arg(A_t),\arg(M_2) < \pi.
\eea
We have fixed the phase on the gluino mass parameter such that 
$\arg(m_{\wt{g}})+\arg(\mu)=\pi$
to maximize the SUSY QCD correction.
When scanning, we imposed the $B(B\to X_s \gamma)$ constraint
and the direct search bounds on the (s)particle masses~\cite{Hagiwara:fs}:
$m_{h^0} \ge 114.3$ GeV and
$m_{\wt{\chi}^\pm_1}, m_{\wt{t}_1}\gtrsim 100$ GeV.

From Fig.~\ref{fig:C8}, we can see that our scenario can easily
accommodate the discrepancy (\ref{eq:sin_b2kp}).
As mentioned above, we have chosen large value for the charged Higgs mass 
$m_{H^\pm}=1$ TeV to 
safely suppress the Barr-Zee type 
two-loop EDM constraints which are significant 
if $\tan\be$~\cite{Chang:1998uc} is large
and the pseudo-scalar Higgs boson ($A^0$) is relatively 
light~\footnote{The SUSY contribution
to $(g-2)_\mu$ is also small in this 
case.~\cite{Arhrib:2001xx}.}. We have checked that for the smaller $m_{H^\pm}$
the larger deviation in $S_{\phi K}$ is possible due to the cancellation
in the real part with the chargino contribution~\cite{baek_long}.
Therefore Fig.~\ref{fig:C8} is a rather conservative result 
for $S_{\phi K}$. We have checked that the allowed range for $S_{\phi K}$
is not sensitive to the change in $\delta_{\rm CKM}$ if we impose the
constraint (\ref{eq:sin2b_exp}).


\begin{figure}
\centering
\includegraphics[width=0.4\textwidth]{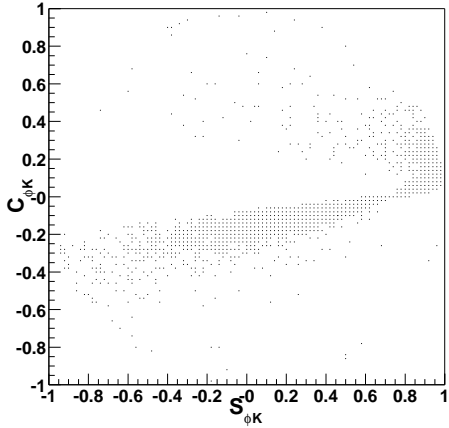}
\caption{Correlation between $C_{\phi K}$ and $S_{\phi K}$
for $\tan\be = 60$.}
\label{fig:C-S}
\end{figure}
$C_{\phi K}$ is correlated with $S_{\phi K}$. In Fig~\ref{fig:C-S},
the correlation  is shown
for $\tan\be = 60$. 
We can see that 
it can also accommodate (\ref{eq:cos_b2kp}) although it favors small
negative $C_{\phi K}$, which will be clarified by future experiments.

The direct CP asymmetry in $B\to X_s \gamma$ decay is also expected to be
large when $S_{\phi K}$ has large deviation from the SM expectations. 
We have checked that actually it can be very large but it is not
correlated with $S_{\phi K}$. It is because $S_{\phi K}$ is determined by
the complex phase on $C_{8g}$ while $A_{CP}(B\to X_s \gamma)$ 
is mainly controlled by that on $C_{7\gamma}$~\cite{Kagan:1998bh}. Because
we have phase on three independent parameters $\mu$, $A_t$ and $M_2$,
the phases of $C_{7\gamma}$ and
$C_{8g}$ need not have strong correlations.

The $B(B\to \phi K)$ varies moderately over the parameter space we considered
and seldom exceeds $15 \times 10^{-6}$, which is acceptable compared with
the experimental measurements~\cite{b2kp_babar,b2kp_belle}. Also the mass
difference $\Delta m_s$ in the $B_s-\ovl{B_s}$ system is close to the
SM expectation $\Delta m_s \sim 14.5$ ps$^{-1}$ in the most region of parameter
space we considered, which may distinguish our scenario from other
scenarios in refs. \cite{b2kp_glu,b2kp:before}.

For large $\tan\be$ in the MSSM, through the Higgs-mediated FCNC the 
$B(B_s \to \mu^+\mu^-)$ can be enhanced by a few orders of magnitude
over the SM prediction~\cite{bsll}. Observation of this leptonic decay mode,
for example at Tevatron Run II,
would be clear signal of NP.
However, this is possible only for relatively light $A^0$.
Since the large deviation in the $S_{\phi K}$,
although it needs new CP violating phase(s) of ${\cal O}(1)$,
does not necessarily require light $A^0$ as we have shown,
these two decay modes can be complementary
to each other in searching for the MSSM at large $\tan\be$.

In our scenario the sole source of large deviation in $S_{\phi K}$ is 
$C_{8g}$. It can have simultaneous effects on 
other decays such as
$B\to \phi\phi$, $B\to \pi\pi$, $B\to K\pi$, {\it etc}~\cite{baek_long}.

In conclusion we have considered the chargino contribution to the CP 
asymmetries $S_{\phi K}$ and $C_{\phi K}$ of $B \to \phi K_S$ decay
in the CP violating MSSM scenario at large $\tan\be$.
We have shown that through the enhanced Wilson coefficient $C_{8g}$ 
of chromo-magnetic penguin operator by the large SUSY threshold 
corrections~\cite{Hall:1993gn}, there can be large deviations in the
CP asymmetries.

\begin{acknowledgments}
The author would like to thank S. Y. Choi for invaluable comments and
carefully reading the manuscript
and E. J. Chun and K. Y. Lee for useful discussions.
\end{acknowledgments}


\begin{thebibliography}{99}

\bibitem{Nir:2002gu}
Y.~Nir,
hep-ph/0208080.


\bibitem{sin2b_babar}
B.~Aubert {\it et al.}  [BABAR Collaboration],
Phys.\ Rev.\ Lett.\  {\bf 87}, 091801 (2001)
[hep-ex/0107013];
B.~Aubert {\it et al.}  [BABAR Collaboration],
Phys.\ Rev.\ Lett.\  {\bf 89}, 201802 (2002)
[hep-ex/0207042].

\bibitem{sin2b_belle}
K.~Abe {\it et al.}  [Belle Collaboration],
Phys.\ Rev.\ Lett.\  {\bf 87}, 091802 (2001)
[hep-ex/0107061];
T.~Higuchi  [Belle Collaboration],
hep-ex/0205020.

\bibitem{b2kp_babar}
B.~Aubert {\it et al.}  [BABAR Collaboration],
hep-ex/0207070.

\bibitem{b2kp_belle}
K.~Abe {\it et al.}  [Belle Collaboration],
hep-ex/0207098.

\bibitem{Hiller:2002ci}
G.~Hiller,
Phys.\ Rev.\ D {\bf 66}, 071502 (2002)
[hep-ph/0207356].



\bibitem{Ciuchini:2002pd}
M.~Ciuchini and L.~Silvestrini,
Phys.\ Rev.\ Lett.\  {\bf 89}, 231802 (2002)
[hep-ph/0208087];


\bibitem{R-parity}
A.~Datta,
Phys.\ Rev.\ D {\bf 66}, 071702 (2002)
[hep-ph/0208016];
B.~Dutta, C.~S.~Kim and S.~Oh,
Phys.\ Rev.\ Lett.\  {\bf 90}, 011801 (2003)
[hep-ph/0208226].



\bibitem{Raidal:2002ph}
M.~Raidal,
Phys.\ Rev.\ Lett.\  {\bf 89}, 231803 (2002)
[hep-ph/0208091];
J.~P.~Lee and K.~Y.~Lee,
hep-ph/0209290.


\bibitem{b2kp_glu}
S.~Khalil and E.~Kou,
hep-ph/0212023;
G.~L.~Kane, P.~Ko, H.~b.~Wang, C.~Kolda, J.~H.~Park and L.~T.~Wang,
hep-ph/0212092;
R.~Harnik, D.~T.~Larson, H.~Murayama and A.~Pierce,
hep-ph/0212180;
M.~Ciuchini, E.~Franco, A.~Masiero and L.~Silvestrini,
hep-ph/0212397.

\bibitem{hep-ex/0301032}
K.~Abe  [Belle Collaboration],
arXiv:hep-ex/0301032.

\bibitem{baek_long}
S.~Baek, work in progress.

\bibitem{b2kp:before}
T.~Moroi,
Phys.\ Lett.\ B {\bf 493}, 366 (2000)
[hep-ph/0007328];
E.~Lunghi and D.~Wyler,
Phys.\ Lett.\ B {\bf 521}, 320 (2001)
[hep-ph/0109149].

\bibitem{Cohen:1996vb}
See, for example, A.~G.~Cohen, D.~B.~Kaplan and A.~E.~Nelson,
Phys.\ Lett.\ B {\bf 388}, 588 (1996)
[arXiv:hep-ph/9607394].



\bibitem{Baek:1998yn}
S.~Baek and P.~Ko,
Phys.\ Rev.\ Lett.\  {\bf 83}, 488 (1999)
[hep-ph/9812229];
S.~Baek and P.~Ko,
Phys.\ Lett.\ B {\bf 462}, 95 (1999)
[hep-ph/9904283].

\bibitem{Beneke:2001ev}
M.~Beneke, G.~Buchalla, M.~Neubert and C.~T.~Sachrajda,
Nucl.\ Phys.\ B {\bf 606}, 245 (2001)
[hep-ph/0104110].

\bibitem{bsr}
N.~Oshimo,
Nucl.\ Phys.\ B {\bf 404}, 20 (1993);
R.~Barbieri and G.~F.~Giudice,
Phys.\ Lett.\ B {\bf 309}, 86 (1993)
[hep-ph/9303270];
P.~L.~Cho, M.~Misiak and D.~Wyler,
Phys.\ Rev.\ D {\bf 54}, 3329 (1996)
[hep-ph/9601360].

\bibitem{Chen:2001fj}
S.~Chen {\it et al.}  [CLEO Collaboration],
Phys.\ Rev.\ Lett.\  {\bf 87}, 251807 (2001)
[arXiv:hep-ex/0108032].

\bibitem{Okada:1990vk}
Y.~Okada, M.~Yamaguchi and T.~Yanagida,
Prog.\ Theor.\ Phys.\  {\bf 85}, 1 (1991).

\bibitem{Hagiwara:fs}
K.~Hagiwara {\it et al.}  [Particle Data Group Collaboration],
Phys.\ Rev.\ D {\bf 66}, 010001 (2002).

\bibitem{Hall:1993gn}
L.~J.~Hall, R.~Rattazzi and U.~Sarid,
Phys.\ Rev.\ D {\bf 50}, 7048 (1994)
[hep-ph/9306309];
T.~Blazek, S.~Raby and S.~Pokorski,
Phys.\ Rev.\ D {\bf 52}, 4151 (1995)
[hep-ph/9504364].

\bibitem{Chang:1998uc}
D.~Chang, W.~Y.~Keung and A.~Pilaftsis,
Phys.\ Rev.\ Lett.\  {\bf 82}, 900 (1999)
[Erratum-ibid.\  {\bf 83}, 3972 (1999)]
[hep-ph/9811202].

\bibitem{Arhrib:2001xx}
A.~Arhrib and S.~Baek,
Phys.\ Rev.\ D {\bf 65}, 075002 (2002)
[hep-ph/0104225];
G.~C.~Cho, N.~Haba and J.~Hisano,
Phys.\ Lett.\ B {\bf 529}, 117 (2002)
[arXiv:hep-ph/0112163];
S.~Baek, P.~Ko and J.~H.~Park,
Eur.\ Phys.\ J.\ C {\bf 24}, 613 (2002)
[arXiv:hep-ph/0203251].


\bibitem{Kagan:1998bh}
A.~L.~Kagan and M.~Neubert,
Phys.\ Rev.\ D {\bf 58}, 094012 (1998)
[hep-ph/9803368].

\bibitem{bsll}
K.~S.~Babu and C.~F.~Kolda,
Phys.\ Rev.\ Lett.\  {\bf 84}, 228 (2000)
[hep-ph/9909476];
C.~S.~Huang, W.~Liao, Q.~S.~Yan and S.~H.~Zhu,
Phys.\ Rev.\ D {\bf 63}, 114021 (2001)
[Erratum-ibid.\ D {\bf 64}, 059902 (2001)]
[hep-ph/0006250];
P.~H.~Chankowski and L.~Slawianowska,
Phys.\ Rev.\ D {\bf 63}, 054012 (2001)
[hep-ph/0008046];
A.~Dedes, H.~K.~Dreiner and U.~Nierste,
Phys.\ Rev.\ Lett.\  {\bf 87}, 251804 (2001)
[hep-ph/0108037];
R.~Arnowitt, B.~Dutta, T.~Kamon and M.~Tanaka,
Phys.\ Lett.\ B {\bf 538}, 121 (2002)
[hep-ph/0203069].
C.~Bobeth, T.~Ewerth, F.~Kruger and J.~Urban,
Phys.\ Rev.\ D {\bf 66}, 074021 (2002)
[hep-ph/0204225].
S.~Baek, P.~Ko and W.~Y.~Song,
Phys.\ Rev.\ Lett.\  {\bf 89}, 271801 (2002)
[hep-ph/0205259];
J.~K.~Mizukoshi, X.~Tata and Y.~Wang,
Phys.\ Rev.\ D {\bf 66}, 115003 (2002)
[hep-ph/0208078].
S.~Baek, P.~Ko and W.~Y.~Song,
hep-ph/0208112;
T.~Ibrahim and P.~Nath,
(to appear in PRD) hep-ph/0208142;
G.~Isidori and A.~Retico,
JHEP {\bf 0209}, 063 (2002)
[hep-ph/0208159];
A.~Dedes and A.~Pilaftsis,
Phys. Rev. D 67, 015012 (2003) [hep-ph/0209306];
S.~Baek, P.~Ko and W.~Y.~Song,
hep-ph/0210373;
A.~J.~Buras, P.~H.~Chankowski, J.~Rosiek and L.~Slawianowska,
hep-ph/0210145.


\end{thebibliography}
\end{document}